\documentclass[aps,showpacs,preprintnumbers,amsmath,amssymb]{revtex4}
\usepackage{dcolumn}
\usepackage[dvips]{epsfig}

\begin{document}
\baselineskip=0.8 cm
\title{\bf Properties of a thin accretion disk around a rotating non-Kerr black hole}

\author{Songbai Chen\footnote{csb3752@163.com}, Jiliang Jing
\footnote{jljing@hunnu.edu.cn}}

\affiliation{Institute of Physics and Department of Physics, Hunan
Normal University,  Changsha, Hunan 410081, People's Republic of
China \\ Key Laboratory of Low Dimensional Quantum Structures \\
and Quantum Control of Ministry of Education, Hunan Normal
University, Changsha, Hunan 410081, People's Republic of China}

\begin{abstract}
\baselineskip=0.6 cm
\begin{center}
{\bf Abstract}
\end{center}

We study the accretion process in the thin disk around a rotating
non-Kerr black hole with a deformed parameter and an unbound
rotation parameter. Our results show that the presence of the
deformed parameter $\epsilon$ modifies the standard properties of
the disk. For the case in which the black hole is more oblate than a
Kerr black hole, the larger deviation leads to the smaller energy
flux, the lower radiation temperature and the fainter spectra
luminosity in the disk. For the black hole with positive deformed
parameter, we find that the effect of the deformed parameter on the
disk becomes more complicated. It depends not only on the rotation
direction of the black hole and the orbit particles, but also on the
sign of the difference between the deformed parameter $\epsilon$ and
a certain critical value $\epsilon_{c}$. These significant features
in the mass accretion process may provide a possibility to test the
no-hair theorem in the strong field regime in future astronomical
observations.

\end{abstract}

 \maketitle
\newpage
\section{Introduction}

It is well known that in general relativity a neutral rotating black
hole in asymptotically flat and matter-free spacetime is described
completely by the Kerr metric only with  two parameters, the mass
$M$ and the rotation parameter $a$, which is supported by the
well-known  no-hair theorem \cite{noh}. This means that all
astrophysical black holes in our Universe should be Kerr black
holes. Although there exists a large amount of observational
evidences for the existence of black holes, a definite proof  is
still lacking at present. In order to test the no-hair theorem,
several potential approaches has been suggested using observations
including gravitational waves from extreme mass-ratio inspirals
\cite{gwave,ga1,ga2,JGa} and the electromagnetic spectrum emitted by
the accreting disk around black holes \cite{ag1,CBa3}, and so on.
However, all of these methods are based on spacetimes that deviate
from the Kerr metric by one or more parameters
\cite{ga2,JGa,ag3,ag4}. The compact object is a Kerr black hole only
if the deviations are measured to be zero.

To test the no-hair theorem, Johannsen \textit{et al} \cite{TJo}
applied recently the Newman-Janis transformation \cite{JNT} and
constructed a deformed Kerr-like black hole metric, which describes
a rotating black hole beyond general relativity. This rotating black
hole possess some striking properties. Besides the mass $M$ and the
rotation parameter $a$, it has a deformation parameter which
measures potential deviations from the Kerr geometry in the
strong-field regime. Especially, there are no restrictions on the
values of the rotation parameter $a$ and the deformation parameter
$\epsilon$. Thus, it is allowable that the rotation parameter is
larger than the mass of the black hole in this alternative gravity
theory, which is different from that in Einstein's general
relativity. Moreover, the radius of horizon of the rotating non-Kerr
black hole depends on the polar angular coordinate $\theta$ and the
spacetime is free of closed timelike curves outside of the outer
horizon \cite{TJo}. If the black hole is more prolate than a Kerr
black hole (i.e., $\epsilon>0$), there exist two disconnected
spherical horizons for high rotation parameter, but there is no
horizon for $a>M$.  If the black hole is more oblate than a Kerr
black hole (i.e., $\epsilon<0$), the horizon always exists for the
arbitrary $a$ and the topology of the horizon becomes toroidal
\cite{CBa,CBa1}. In the weak field approximation, such a black hole
possesses the same asymptotic behaviors of the usual Kerr black hole
in general relativity \cite{TJo}. Motivated by testing gravity in
the strong field region, a lot of efforts has been recently
dedicated to the study of the rotating non-Kerr black hole
\cite{FCa,TJo,CBa, CBa1, VCa1,TJo1}.

It is well known that the accretion processes is a powerful
indicator of the physical nature of the central celestial objects,
which means that the analysis of the signatures of the accretion
disk around the rotating non-Kerr black hole could help us to detect
gravity effects in the strong field regime in which general
relativity breaks down. The accretion disk is such a structure
formed by the diffuse material in orbital motion around a central
compact body, which now is an important research topic in the
astrophysics.  The steady-state thin accretion disk model is the
simplest theoretical model of the accretion disks, in which the disk
has negligible thickness so that the heat generated by stress and
dynamic friction in the disk can be dispersed through the radiation
over its surface \cite{sdk1,sdk2,Page,Thorne}. This cooling
mechanism ensures that the disk can be in hydrodynamical equilibrium
and the mass accretion rate in the disk maintains a constant, which
is independent of time variable. The physical properties of matter
forming a thin accretion disk in a variety of background spacetimes
have been investigated extensively in
\cite{Harko,Bhattacharyya,Kovacs,Torres,Yuan,Guzman,Pun,Cs,ZLP,CBa3,CBa4,JGa}.
The special signatures appeared in the energy flux and the emission
spectrum emitted by the disk can provide us not only the information
about black holes in the Universe, but also the profound
verification of alternative theories of gravity. The marginally
stable orbit for the particle and the conversion efficiency of the
central object converting rest mass into outgoing radiation  have
been studied in the background of a rotating non-Kerr black hole
\cite{TJo,CBa1}, which tells us that with the increase of the
deformed parameter $\epsilon$, the marginally stable orbit radius
$r_{ms}$ decreases and the conversion efficiency increases. However,
the effects of the deformation parameter on the energy flux, the
radiation temperature, the spectra luminosity and the spectra
cut-off frequency is still open. The main purpose of the present
Letter is to study the properties of the thin accretion disk in the
rotating non-Kerr black hole spacetime and see whether it can leave
us the signature of the deformation parameter in the energy flux and
the emission spectrum emitted in the mass accretion process.

The Letter is organized as follows: in the following section we will
review briefly the rotating no-Kerr black hole metric proposed by
Johannsen \textit{et al} \cite{TJo} to test the no-hair theorem in
the strong field regime, and then present the geodesic equations for
the timelike particles moving in the equatorial plane in this
background. In Sec.III, we study the physical properties of the thin
accretion disk around the rotating no-Kerr black hole  and probe the
effects of the deformation parameter on the energy flux, temperature
and emission spectrum of the thin accretion disks onto this black
hole. We end the Letter with a summary.

\section{The geodesic equations in the rotating non-Kerr black hole spacetime}

Let us now review briefly the rotating no-Kerr black hole metric,
which is proposed by Johannsen \textit{et al} \cite{TJo} to test
gravity in the strong field regime. Starting from a deformed
Schwarzschild solution and applying the Newman-Janis transformation,
they constructed a deformed Kerr-like metric which describes a
stationary, axisymmetric, and asymptotically flat spacetime. It
contains three parameters: the mass $M$, the rotation parameter $a$,
and the deformation parameter $\epsilon$. In the standard
Boyer-Lindquist coordinates, the metric of this rotating no-Kerr
black hole has a form \cite{TJo}
\begin{eqnarray}
ds^2=g_{tt}dt^2+g_{rr}dr^2+g_{\theta\theta}d\theta^2+g_{\phi\phi}
d\phi^2+2g_{t\phi}dtd\phi, \label{metric0}
\end{eqnarray}
where
\begin{eqnarray}
g_{tt}&=&-\bigg(1-\frac{2Mr}{\rho^2}\bigg)(1+h),\;\;\;\;\;
g_{t\phi}=-\frac{2aMr\sin^2\theta}{\rho^2}(1+h),\nonumber\\
g_{rr}&=&\frac{\rho^2(1+h)}{\Delta+a^2h\sin^2\theta},\;\;\;\;\;\;\;\;\;\;\;\;\;\;\;
g_{\theta\theta}=\rho^2,\nonumber\\
g_{\phi\phi}&=&\sin^2\theta\bigg[r^2+a^2+\frac{2a^2Mr\sin^2\theta}{\rho^2}\bigg]
+\frac{a^2(\rho^2+2Mr)\sin^4\theta}{\rho^2}h,
\end{eqnarray}
with
\begin{eqnarray}
\rho^2=r^2+a^2\cos^2\theta,\;\;\;\;\;\;\;\;\;\;
\Delta=r^2-2Mr+a^2,\;\;\;\;\;\;\;\;\;\;h=\frac{\epsilon M^3
r}{\rho^4}.
\end{eqnarray}
The deformed parameter $\epsilon>0$ or $\epsilon<0$ corresponds to
the cases in which the compact object is more prolate or oblate than
a Kerr black hole, respectively. As $\epsilon=0$, the black hole is
reduced to the usual Kerr black hole in general relativity. The
horizon of the black hole is given by \cite{CBa,CBa1}
\begin{eqnarray}
\Delta+a^2h\sin^2\theta=0.
\end{eqnarray}
Clearly, the radius of horizon depends on $\theta$, which is
different from that in the usual Kerr case. For the case
$\epsilon>0$, one can find that there exist two disconnected
spherical horizons for high spin parameters, but there is no horizon
for $a>M$. However, for $\epsilon<0$ one can find that the horizon
never disappears for the arbitrary $a$ and the shape of the horizon
becomes toroidal \cite{CBa,CBa1}. Moreover, one can find that it is
free of closed timelike curves and then causality is satisfied
outside of the event horizon \cite{TJo}.

According to the thin accretion disk model, one can assumes that the
disk is on the equatorial plane and that the matter moves on nearly
geodesic circular orbits. In the rotating non-Kerr black hole
spacetime (\ref{metric0}), the timelike geodesics equations of a
particle can be expressed as
\begin{eqnarray}
&&u^{t}=\frac{dt}{d\lambda}=\frac{\tilde{E}g_{\phi\phi}-\tilde{L}g_{t\phi}}{g^2_{t\phi}-g_{tt}g_{\phi\phi}},\label{u1}\\
&&u^{\phi}=\frac{d\phi}{d\lambda}=\frac{\tilde{E}g_{t\phi}+\tilde{L}g_{tt}}{g^2_{t\phi}-g_{tt}g_{\phi\phi}},\label{u2}\\
&&g_{rr}\bigg(\frac{dr}{d\lambda}\bigg)^2+g_{\theta\theta}\bigg(\frac{d\theta}{d\lambda}\bigg)^2=V_{eff},
\end{eqnarray}
with the effective potential
\begin{eqnarray}
V_{eff}=\frac{\tilde{E}^2g_{\phi\phi}+2\tilde{E}\tilde{L}g_{t\phi}+\tilde{L}^2g_{tt}}{g^2_{t\phi}-g_{tt}g_{\phi\phi}}-1,
\end{eqnarray}
where $\tilde{E}$ and $\tilde{L}$ are the specific energy and the
specific angular momentum of the particle, respectively.

The circular equatorial orbits obey the conditions $V_{eff}=0$,
$V_{eff,r}=0$ and $V_{eff,\theta} =0$ \cite{CBa3,CBa4,JGa}. Due to
the refection symmetry of the metric (\ref{metric0}) with respect to
the equatorial plane, one can find that the condition
$V_{eff,\theta} =0$ is satisfied naturally for the particles
locating at the plane $\theta=\pi/2$. Making use of these
conditions, we can get the specific energy $\tilde{E}$, the specific
angular momentum $\tilde{L}$, and the angular velocity $\Omega$ of
the particle moving in circular orbit on the equatorial plane in the
rotating non-Kerr black hole spacetime
\begin{eqnarray}
&&\tilde{E}=-\frac{g_{tt}+g_{t\phi}\Omega}{\sqrt{g_{tt}-2g_{t\phi}\Omega-g_{\phi\phi}\Omega^2}}
\label{sE}\\
&&\tilde{L}=\frac{g_{t\phi}+g_{\phi\phi}\Omega}{\sqrt{g_{tt}-2g_{t\phi}\Omega-g_{\phi\phi}\Omega^2}}\;, \label{sL}\\
&&\Omega=\frac{d\phi}{dt}=\frac{-g_{t\phi,r}+\sqrt{g_{t\phi,r}^2-g_{tt,r}g_{\phi\phi,r}}}{g_{\phi\phi,r}}.
\end{eqnarray}
The circular orbits are stable under small perturbations in the
radial direction if $V_{eff ,rr}\leq0$, and in the vertical
direction if $V_{eff ,\theta\theta}\leq0$ \cite{CBa3,CBa4,JGa}. In
the background of a Kerr black hole in general relativity, the
second condition is always satisfied, so one can obtain the radius
of the innermost stable circular orbit by solving $V_{eff ,rr}=0$.
However, in the rotating black hole beyond general relativity, the
marginally stable orbit radius is determined by the condition
$V_{eff ,rr}=0$ or $V_{eff ,\theta\theta}=0$ \cite{CBa3,CBa4,JGa}.
Unfortunately, in a rotating non-Kerr black hole (\ref{metric0}),
one can not obtain an analytical form of the marginally stable orbit
radius from these conditions. In order to get the marginally stable
orbit radius, we must resort to numerical method. In Fig.(1), we set
$M=1$ and plotted the variety of the marginally stable orbit radius
$r_{ms}$ with the deformed parameter $\epsilon$ in the rotating
non-Kerr black hole spacetime. When the black hole is oblate than a
Kerr black hole ($\epsilon<0$), one can find that the radius
$r_{ms}$ decreases with the deformed parameter $\epsilon$ and the
rotation parameter $a$, which is similar that in the Kerr black hole
spacetime. This is also shown in \cite{TJo}. When the black hole is
prolate ($\epsilon>0$), the dependence of the marginally stable
orbit radius $r_{ms}$ on the deformed parameter $\epsilon$ becomes
more complicated. Let us now first to discuss the case where the
particle rotates in the same direction to this prolate black hole
($a>0$). As the parameter $\epsilon<\epsilon_{c1}$, the marginally
stable orbit radius $r_{ms}$ is defined by the radial instability as
in the oblate black hole case and $r_{ms}$  decreases with the
deformed parameter $\epsilon$.
\begin{figure}[ht]
\begin{center}
\includegraphics[width=8cm]{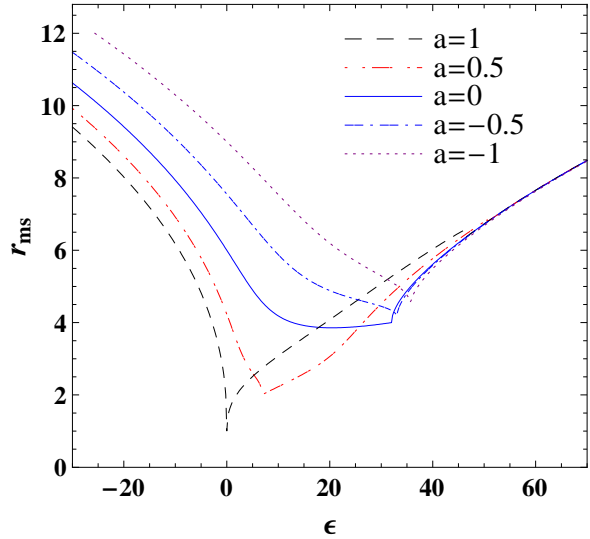}
\caption{Variety of the marginally stable orbit radius $r_{ms}$ with
the parameter $\epsilon$ for the thin disk around the
rapidly-rotating non-Kerr black hole.}
\end{center}
\end{figure}
\begin{table}[h]
\begin{center}
\begin{tabular}{c|ccccc}
\hline\hline  &&&&&\\
$a$&0&0.3&0.5&0.8&1.0 \\
\hline
&&&&&\\
$\epsilon_{c1}$& 31.9&28.9&6.9&0.91&0 \\\hline
&&&&&\\
$\epsilon_{c2}$& $\infty$&51.61&50.73&48.54&46.51 \\
\hline\hline &&&&&\\
$a$&-0.3&-0.5&-0.8&-0.9&-1.0 \\
\hline
&&&&&\\
$\epsilon_{c3}$& 32.40&33.06&34.50&35.07&35.67 \\
\hline\hline
\end{tabular}
\end{center}
\label{tab1} \caption{The change of the critical values
$\epsilon_{c1}$, $\epsilon_{c2}$,
 and $\epsilon_{c3}$ with the rotation parameter $a$. Here we set $M=1$.}
\end{table}
As $\epsilon_{c1}<\epsilon<\epsilon_{c2}$, there are two
disconnected regimes where stable circular orbits exist: an outer
zone with $r>r_1$ and an inner zone with $r_3<r<r_2$. Since the
energy and angular momentum of the orbit in the inner zone are
higher than that in the outer zone, an object inspiraling from large
distances on a circular equatorial orbit will reach $r_1$ and plunge
into the central body, rather than finding itself in the inner range
of circular orbits. This means that the marginally stable orbit
radius $r_{ms}$ is consistent with $r_1$, which is defined by the
vertical instability. One can see that in this regime the marginally
stable orbit radius $r_{ms}$ increases with $\epsilon$. Those
properties of $r_{ms}$ are similar to those of in the Manko-Novikov
metric \cite{CBa4,JGa}. As $\epsilon>\epsilon_{c2}$, the outer zone
is connected with the inner zone. However, the energy and angular
momentum of the orbit are positive and real only if $r>r_4$. Thus,
the effective marginally stable orbit is at $r_4$ and increases with
$\epsilon$, which is different from that in the background of a
Manko-Novikov spacetime \cite{CBa4,JGa}. Moreover, we also find that
the effective marginally stable orbit radius $r_4$ is independent of
$a$ for the larger $\epsilon$. The critical values of
$\epsilon_{c1}$ and $\epsilon_{c2}$ are listed in the Table (1) for
different rotation parameter $a$. It is shown that both
$\epsilon_{1c}$ and $\epsilon_{2c}$ decrease with $a$. For the
particle rotating in the opposite direction to the black hole
($a<0$), we find as the deformed parameter $\epsilon$ is less than a
certain critical value $\epsilon_{c3}$, the marginally stable orbit
radius is defined by the radial instability as in the previous
discussion, which $r_{ms}$ decreases monotonously with $\epsilon$.
As $\epsilon>\epsilon_{c3}$, we find that only in the region where
the energy and angular momentum of the orbit are positive and real,
the orbit is stable under small perturbations both in the radial
direction and in the vertical direction since  both of the
conditions $V_{eff ,rr}\leq0$ and $V_{eff ,\theta\theta}\leq0$ are
satisfied. Thus, the effective marginally stable orbit is still at
$r_4$. The upper bound $\epsilon_{c3}$ increases with the absolute
value of $a$ as is shown in Table (1). Moreover, we note that for
the value of $\epsilon$ is much larger than $\epsilon_{c2}$ for
$a>0$ and than $\epsilon_{c3}$ for $a<0$, the marginally stable
orbit radius is independent of the rotation parameter $a$.

\section{The properties of thin accretion disks in the
rotating non-Kerr black hole spacetime}

In this section we will adopt the steady-state thin accretion disk
model to study the accretion process in the thin disk around the
rotating non-Kerr black hole and probe how the deformed parameter
$\epsilon$ affects the energy flux, the conversion efficiency, the
radiation temperature and the spectra of the disk in this
background.  In the steady-state accretion disk models, the
accreting matter in the disk can be described by an anisotropic
fluid with the energy-momentum tensor \cite{Page,Thorne}
\begin{eqnarray}
T^{\mu\nu}=\varepsilon_0
u^{\mu}u^{\nu}+2u^{(\mu}q^{\nu)}+t^{\mu\nu},
\end{eqnarray}
where the quantities $\varepsilon_0$, $q^{\mu}$ and $t^{\mu\nu}$
denotes the rest mass density, the energy flow vector and the stress
tensor of the accreting matter, respectively,  which are defined in
the averaged rest-frame of the orbiting particle with four-velocity
$u^{\mu}$. In the averaged rest-frame, we have $u_{\mu}q^{\mu}=0$
and $u_{\mu}t^{\mu\nu}=0$ since both $q^{\mu}$ and $t^{\mu\nu}$ is
orthogonal to $u^{\mu}$ \cite{Page,Thorne}.
\begin{figure}[ht]
\begin{center}
\includegraphics[width=5.4cm]{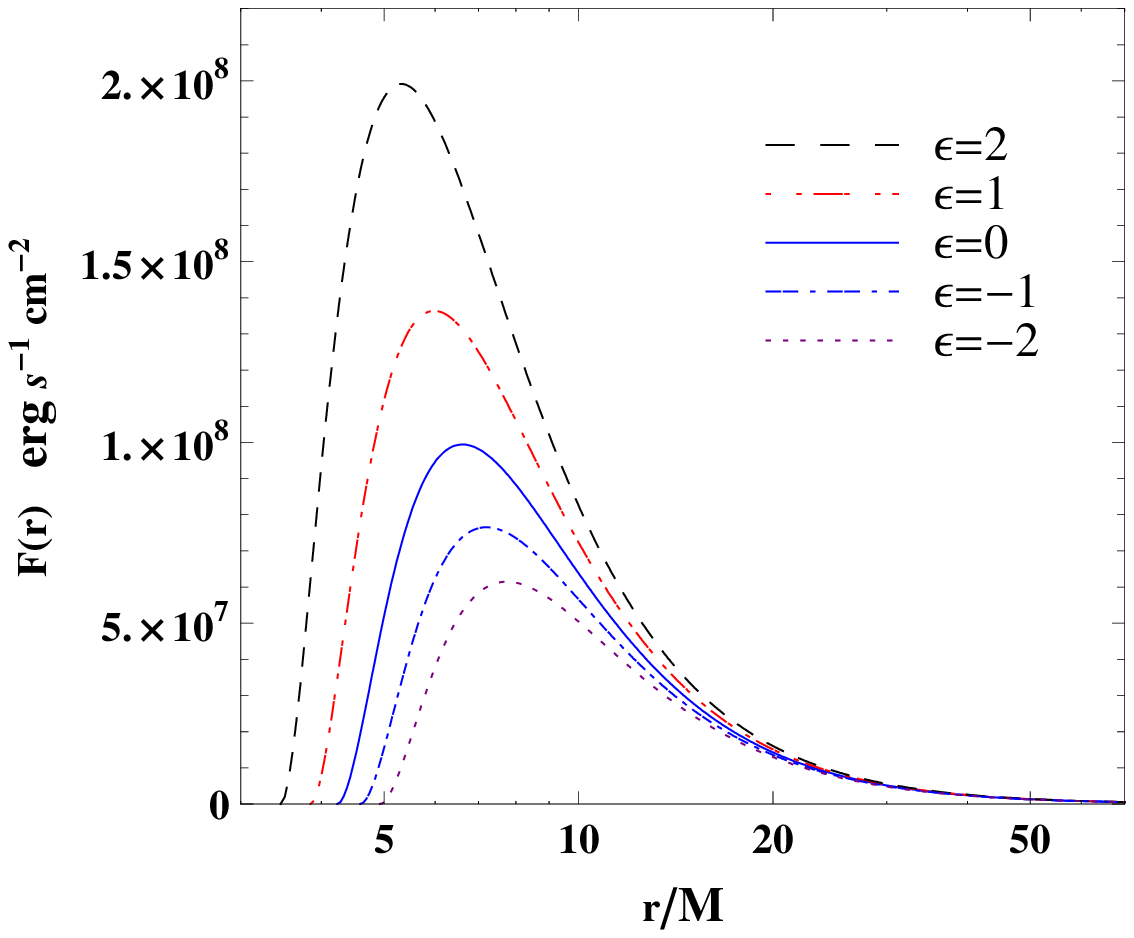}\;\;\includegraphics[width=5.4cm]{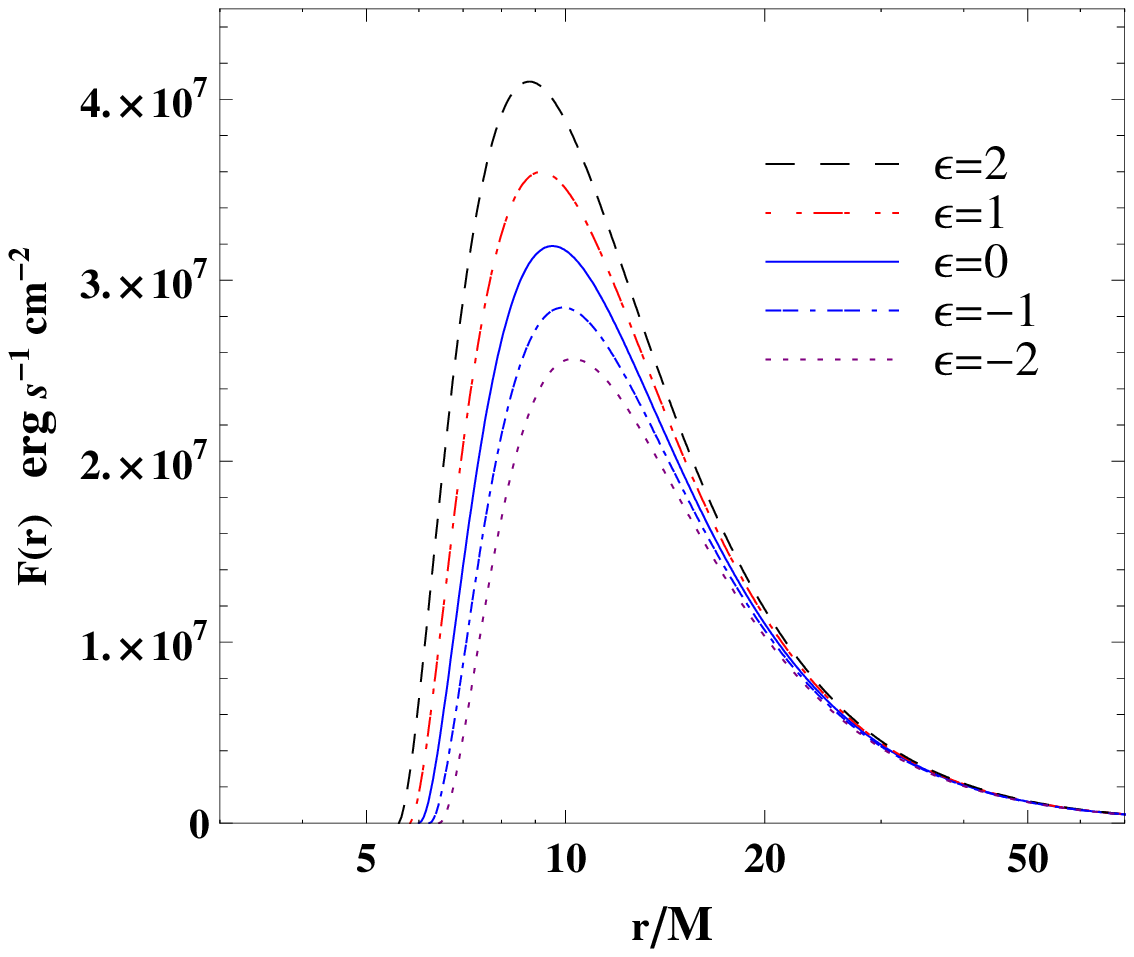}
\;\includegraphics[width=5.4cm]{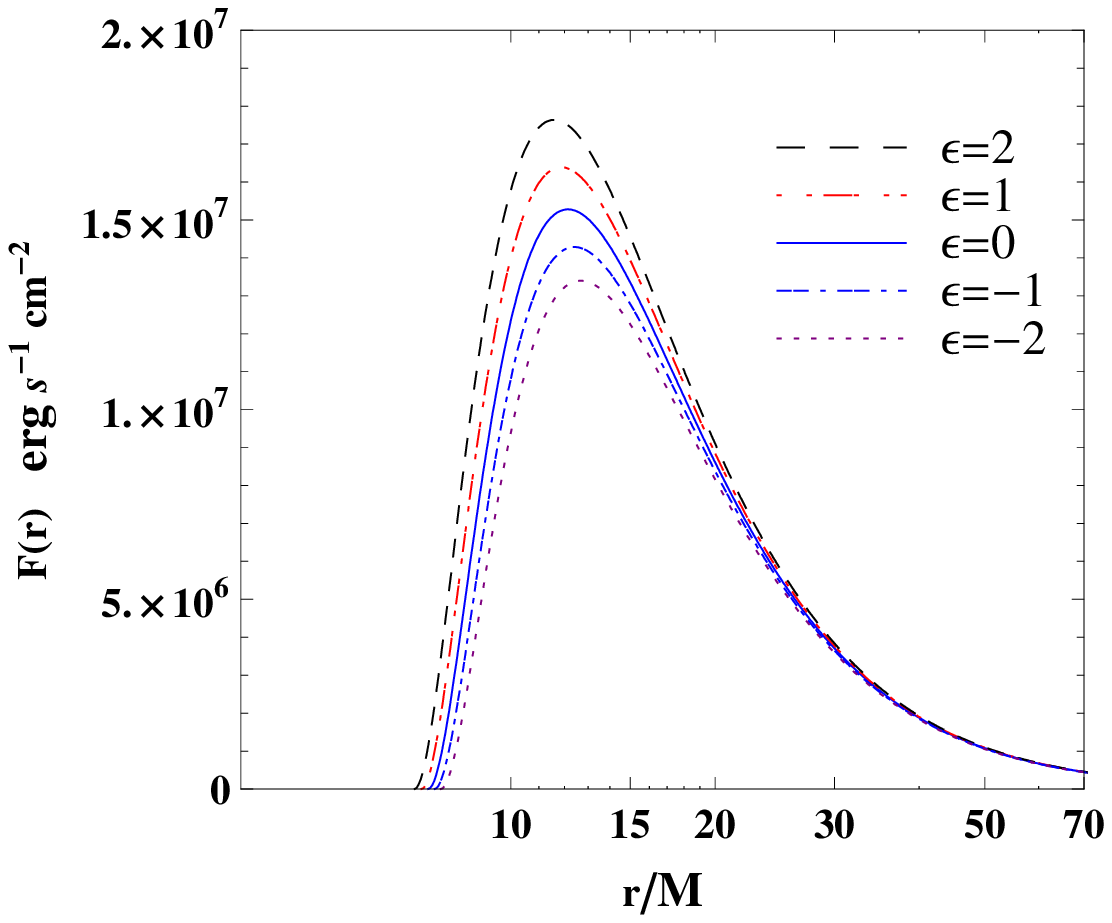}\\\includegraphics[width=5.4cm]{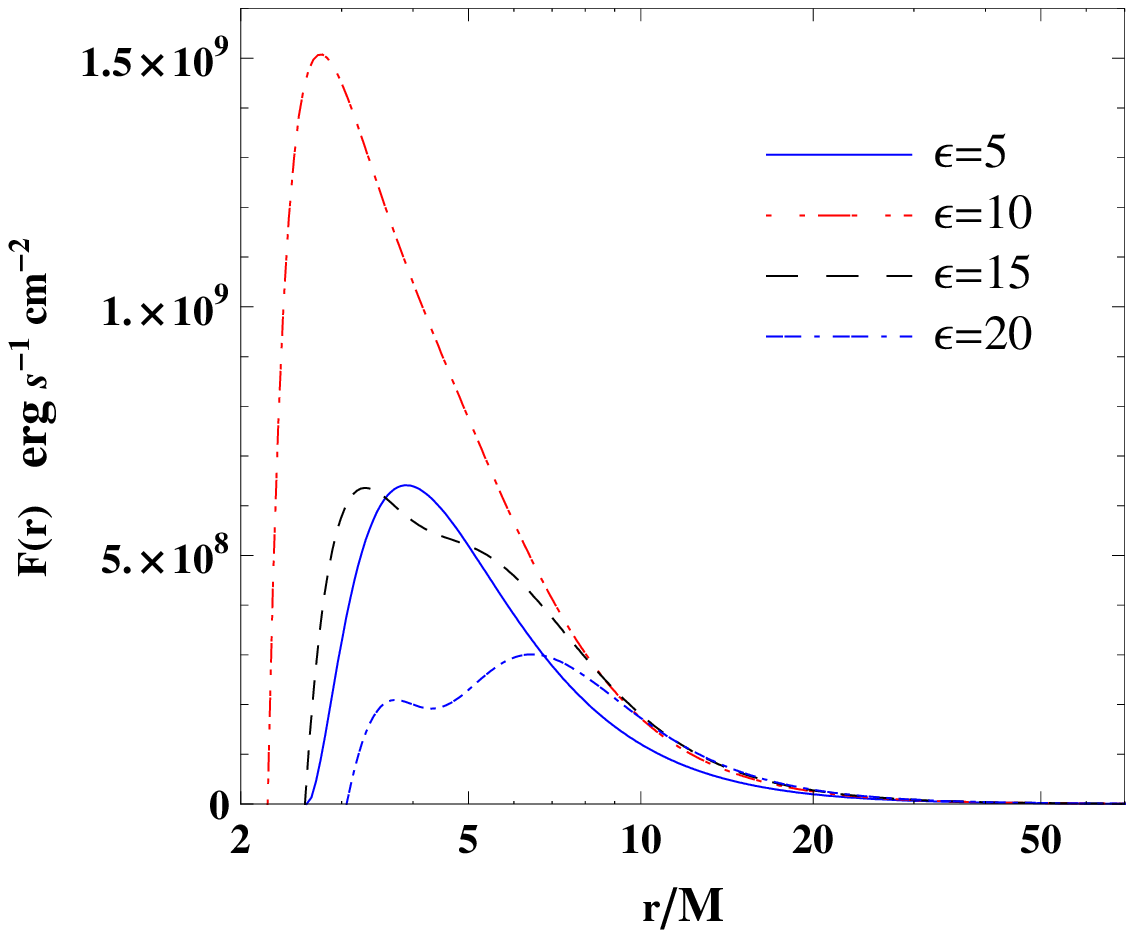}
\;\;\includegraphics[width=5.4cm]{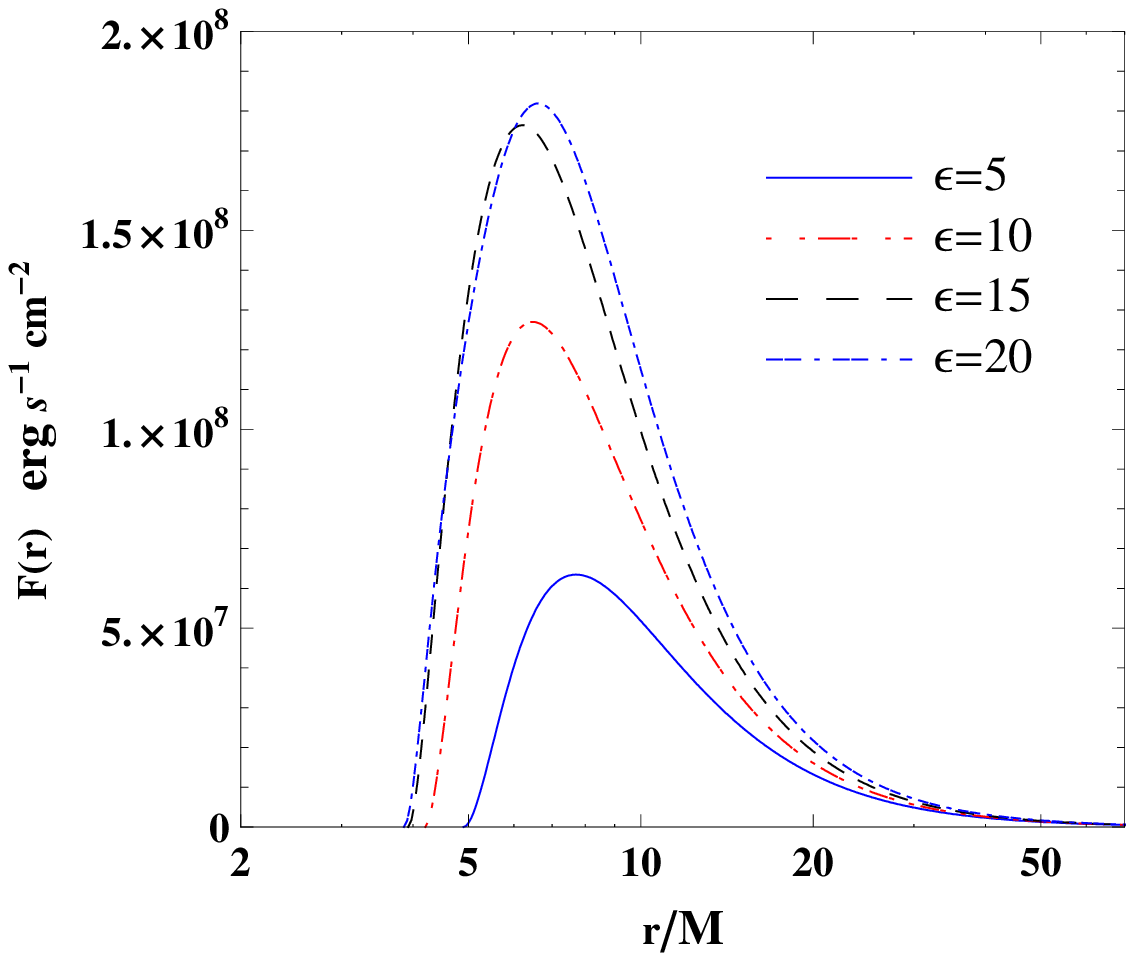}
\;\includegraphics[width=5.4cm]{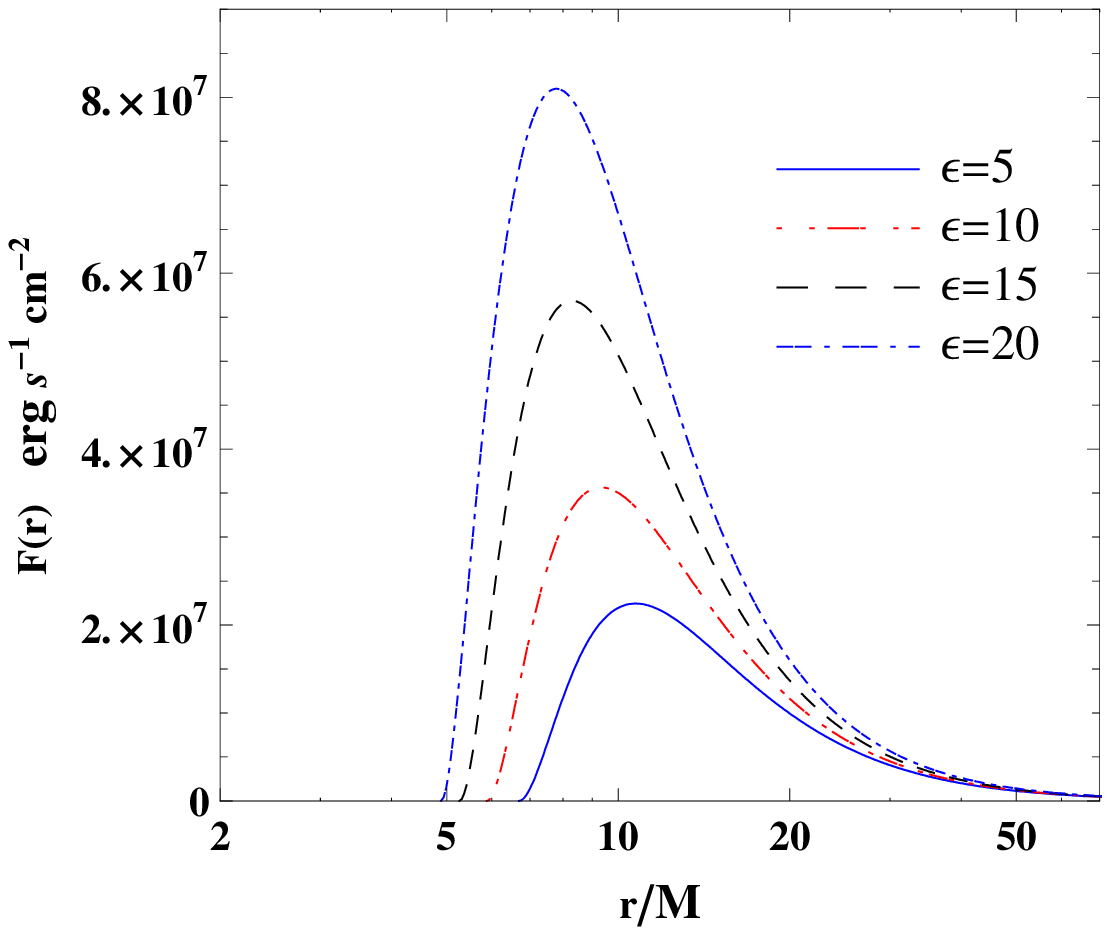} \caption{Variety of
the energy flux $F(r)$ with the deformed parameter $\epsilon$ in the
thin disk around the rotating non-Kerr black hole. The two panels at
the left, middle and right correspond to the cases when $a=0.5$,
$a=0$ and $a=-0.5$, respectively. Here, we set the total mass of the
black hole $M=10^6M_{\odot}$ and the mass accretion rate
$\dot{M_0}=10^{-12}M_{\odot}\;yr^{-1}$.}
\end{center}
\end{figure}

In background of the rotating non-Kerr black hole, one can find that
the time-averaged radial structure equations of the thin disk can be
expressed as
\begin{eqnarray}
&&\dot{M_0}=-2\pi\sqrt{-G}\Sigma(r) u^{r}=\text{Const},\\
&&\bigg[\dot{M_0}\tilde{E}- 2\pi\sqrt{-G}\Omega
W_{\phi}^{\;r}\bigg]_{,r}
=2\pi\sqrt{-G}F(r)\tilde{E},\label{Ws1}\\
&&\bigg[\dot{M_0}\tilde{L}-2\pi\sqrt{-G}W_{\phi}^{\;r}\bigg]_{,r}
=2\pi\sqrt{-G}F(r)\tilde{L},\label{Ws2}
\end{eqnarray}
with
\begin{eqnarray}
\Sigma(r)=\int^{H}_{-H}\langle\varepsilon_0\rangle
\;dz,\;\;\;\;\;\;\; W_{\phi}^{\;r}=\int^{H}_{-H}\langle
t_{\phi}^{\;r} \rangle
\;dz,\;\;\;\;\;\;\;\sqrt{-G}=\frac{(r^3+\epsilon)}{r^2},
\end{eqnarray}
where $\Sigma(r)$ and  $W_{\phi}^{\;r}$ are the averaged rest mass
density and the averaged torque, respectively. The quantity $\langle
t_{\phi}^{\;r}\rangle$ is the average value of the $\phi-r$
component of the stress tensor over a characteristic time scale
$\Delta t$ and the azimuthal angle $\Delta\phi=2\pi$. With the
energy-angular momentum relation for circular geodesic orbits
$\tilde{E}_{,r}=\Omega \tilde{L}_{,r}$, one can eliminate
$W_{\phi}^{\;r}$ from Eqs.(\ref{Ws1}) and (\ref{Ws2}), and then
obtain the expression of the energy flux in the mass accretion
process
\begin{eqnarray}
F(r)=-\frac{\dot{M_0}}{4\pi\sqrt{-G}}
\frac{\Omega_{,r}}{(\tilde{E}-\Omega\tilde{L})^2}\int^{r}_{r_{ms}}
(\tilde{E}-\Omega\tilde{L})\Omega_{,r}dr.\label{enf}
\end{eqnarray}

Here, we consider the mass accretion driven by black holes with a
total mass of $M=10^6M_{\odot}$, and with a mass accretion rate of
$\dot{M_0}=10^{-12}M_{\odot}\;yr^{-1}$ \cite{Harko}. In Fig. (2), we
plotted the total energy flux $F(r)$ radiated by a thin disk around
the rotating non-Kerr black hole for different deformed parameter
$\epsilon$ and rotation parameter $a$. When the black hole is oblate
than a Kerr black hole ($\epsilon<0$), one can find that the energy
flux $F(r)$ increases with the deformed parameter $\epsilon$ and the
rotation parameter $a$, which is similar to that in the Kerr black
hole spacetime. When the black hole is prolate ($\epsilon>0$), the
situation becomes more complex. As $a<0$, the flux $F(r)$ increases
with $\epsilon$ for $\epsilon<\epsilon_{c3}$ and decreases with
$\epsilon$  for $\epsilon>\epsilon_{c3}$. As $a>0$, we find that
with the increase of $\epsilon$, $F(r)$ increases as
$\epsilon<\epsilon_{c1}$ and decreases as $\epsilon>\epsilon_{c1}$.
Comparing with the behavior of $r_{ms}$, one can obtain that the
dependence of $F(r)$ on $\epsilon$ is converse to those of $r_{ms}$,
which can be explained by a fact that the larger $r_{ms}$ enhances
the lower limit of integral in the energy flux (\ref{enf}). With the
increase of $\epsilon$, the position of the peak value of $F(r)$
moves along the left for the smaller $\epsilon$ and moves along
right for the larger $\epsilon$. When $\epsilon>\epsilon_{c1}$, the
multi-peak value appears in the energy flux $F(r)$. Moreover, we
also find that the difference in the energy flux originating from
the parameter $\epsilon$ increases with the rotation parameter $a$
as $\epsilon<\epsilon_{c2}$ for $a>0$ and $\epsilon<\epsilon_{c3}$
for $a<0$. This means that in this region as the black hole rotates
more rapidly, the effect of the deformed parameter $\epsilon$ on the
energy flux $F(r)$ becomes more distinct for the prograde particles
and more tiny for the retrograde ones. When $\epsilon>\epsilon_{c2}$
for $a>0$ and $\epsilon>\epsilon_{c3}$ for $a<0$, the difference in
the energy flux originating from the rotation parameter $a$ vanishes
gradually with the increase of $\epsilon$ because $r_{ms}$ does not
depend on $a$ for the larger $\epsilon$ in this region.
\begin{figure}[ht]
\begin{center}
\includegraphics[width=7cm]{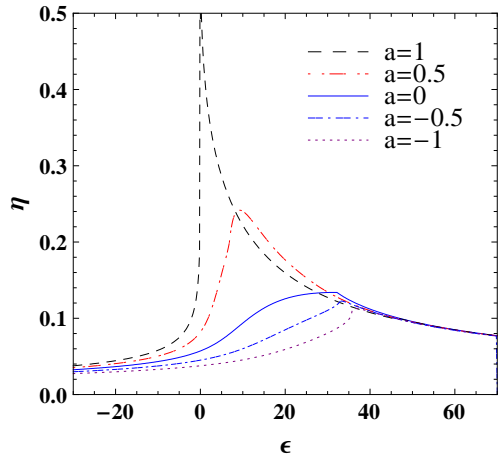}
\caption{Variety of the efficient $\eta$ with the parameter
$\epsilon$ for the thin disk around the rotating non-Kerr black
hole.}
\end{center}
\end{figure}

In the mass accretion process around a black hole, the conversion
efficiency is an important characteristic quantity which describes
the capability of the central object converting rest mass into
outgoing radiation. In general, the conversion efficiency can be
given by the ratio of two rates measured at infinity
\cite{sdk2,Page}: the rate of the radiation energy of photons
escaping from the disk surface to infinity and the mass-energy
transfer rate of the central compact object in the mass accretion.
If all the emitted photons can escape to infinity, one can find that
the efficiency $\eta$ is determined by the specific energy of a
particle at the marginally stable orbit $r_{ms}$\cite{Thorne}
\begin{eqnarray}
\eta=1-\tilde{E}_{ms}.\label{effi}
\end{eqnarray}
The dependence of the conversion efficiency $\eta$ on the deformed
parameter $\epsilon$ is plotted in Fig.(3), which shows that when
$\epsilon<\epsilon_{c3}$ for $a<0$ and $\epsilon<\epsilon_{c1}$ for
$a>0$, the larger values of $\epsilon$ leads to a much larger
efficiency $\eta$. This means that in this parameter region the
conversion efficiency of the thin accretion disk in the prolate
non-Kerr black hole spacetime is more than that in the oblate one.
However, as $\epsilon>\epsilon_{c3}$ for $a<0$ and
$\epsilon>\epsilon_{c1}$ for $a>0$, we find that efficiency $\eta$
decreases with $\epsilon$. Similarly, for the larger positive
$\epsilon$, the conversion efficiency is dominated by the deformed
parameter $\epsilon$ and independent of the rotation parameter $a$
of the central object. The relationship between the rotation
parameter $a$ and deformation parameter $\epsilon$ for fixed
conversion efficiency $\eta$ has been discussed in \cite{CBa1}.
\begin{figure}[ht]
\begin{center}
\includegraphics[width=5.4cm]{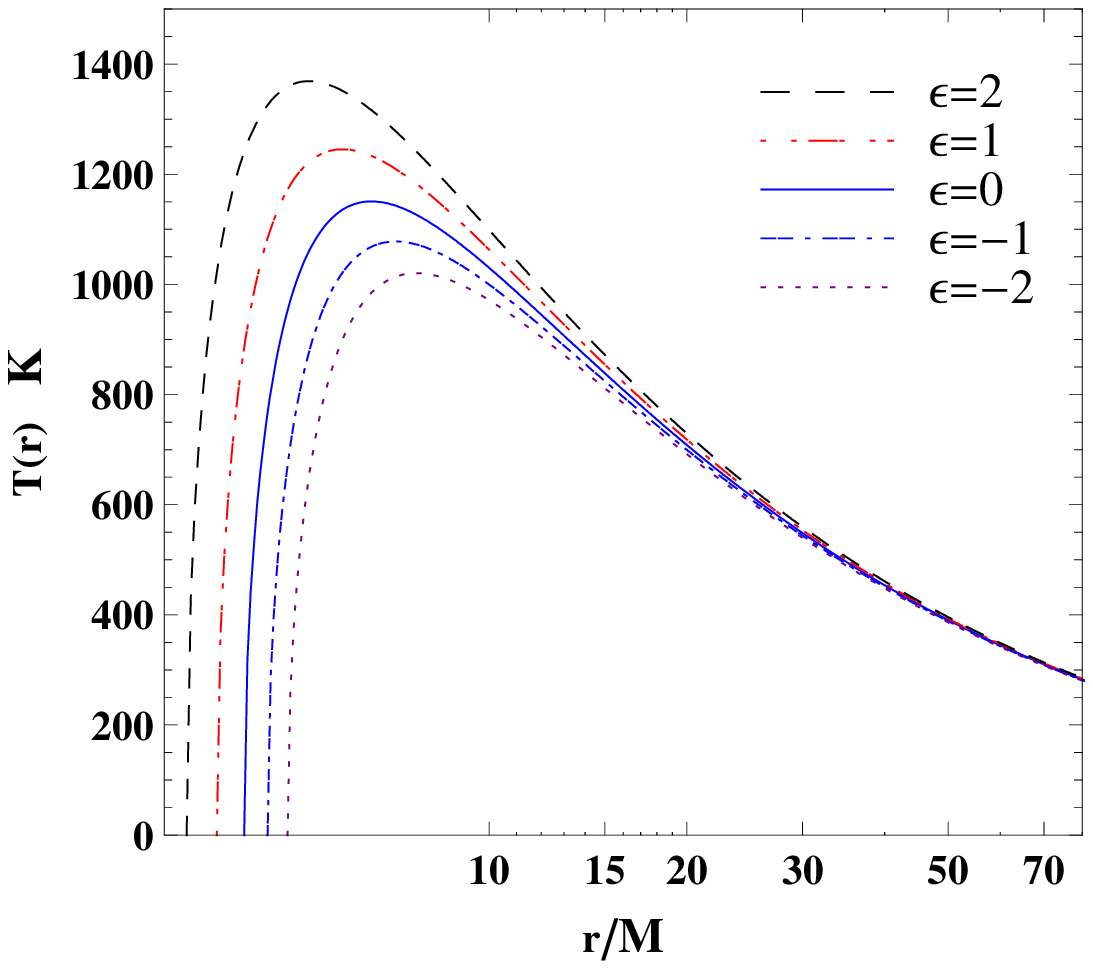}\includegraphics[width=5.4cm]{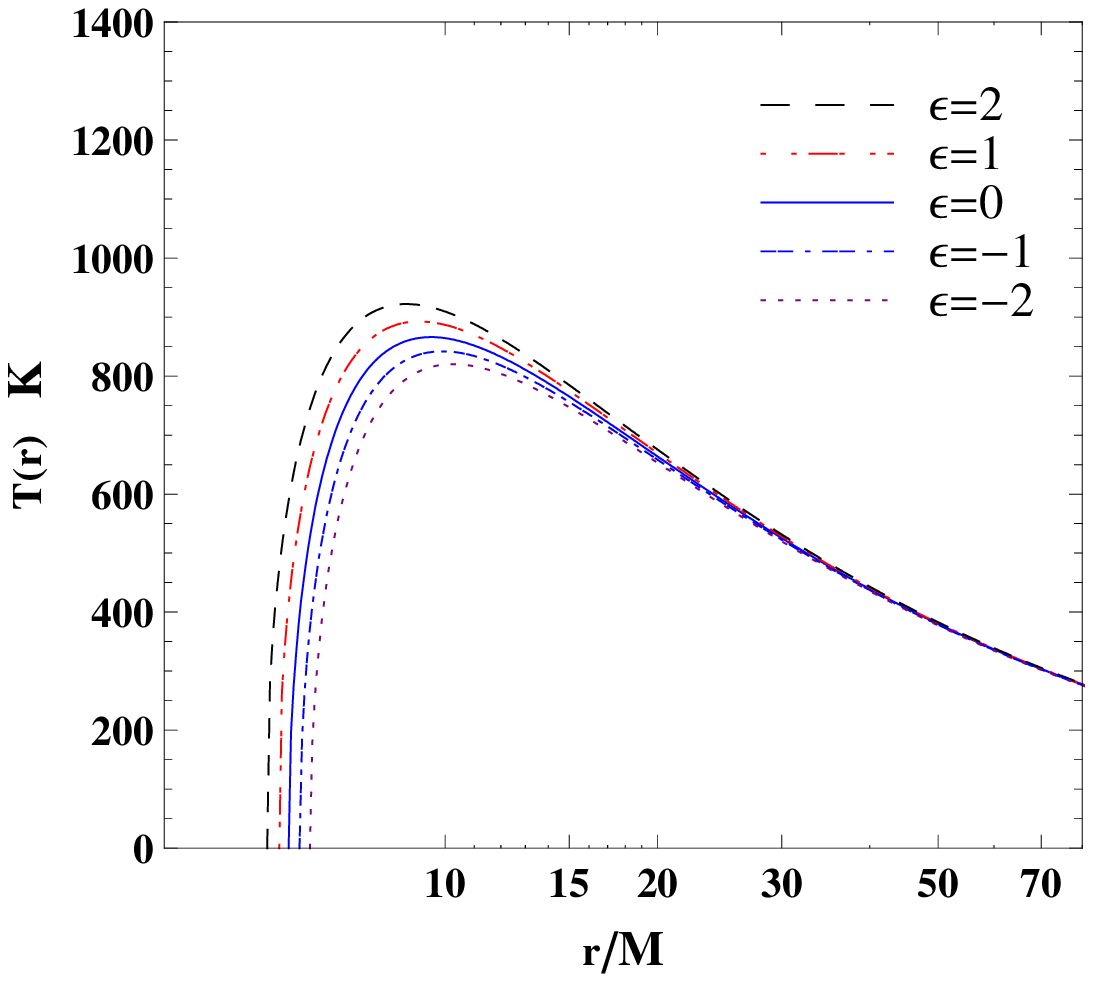}
\includegraphics[width=5.4cm]{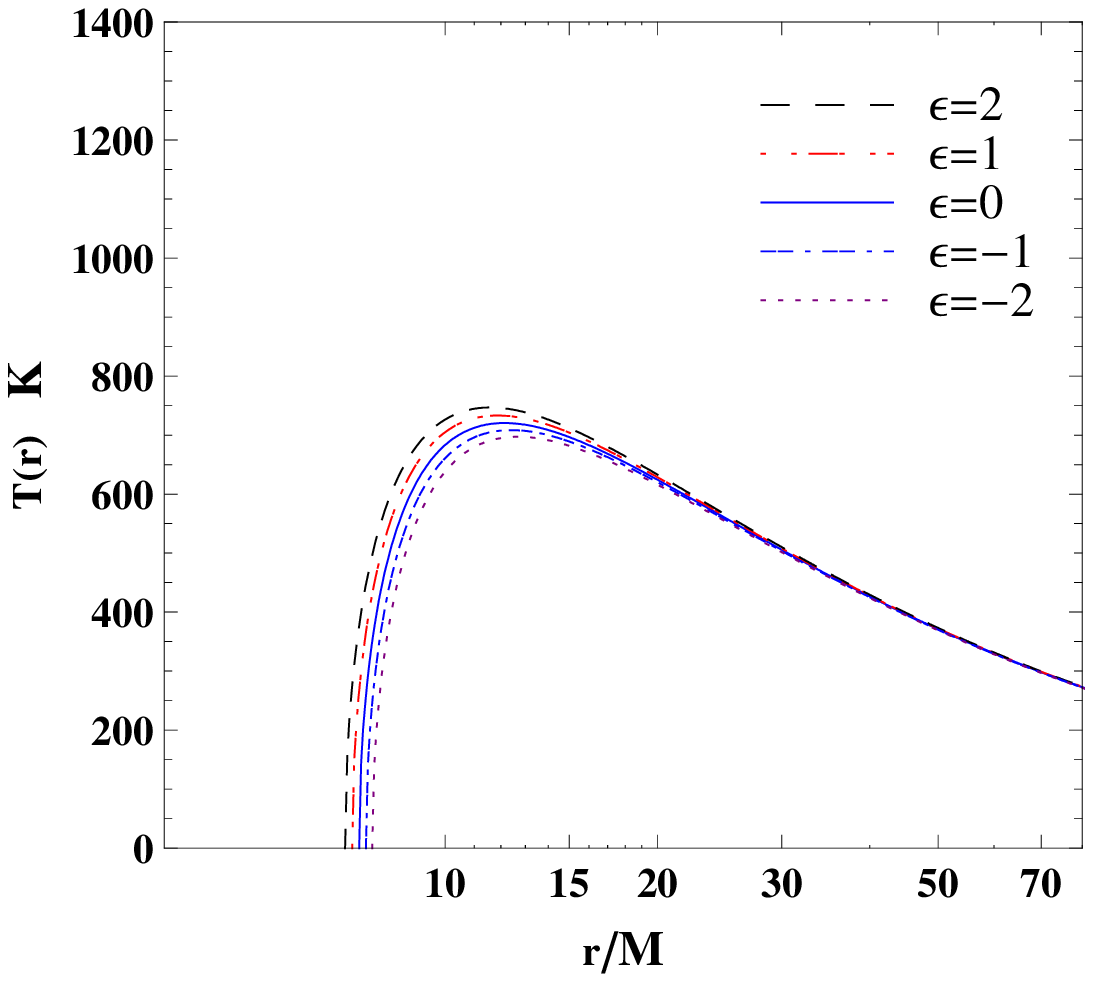}\\
\includegraphics[width=5.4cm]{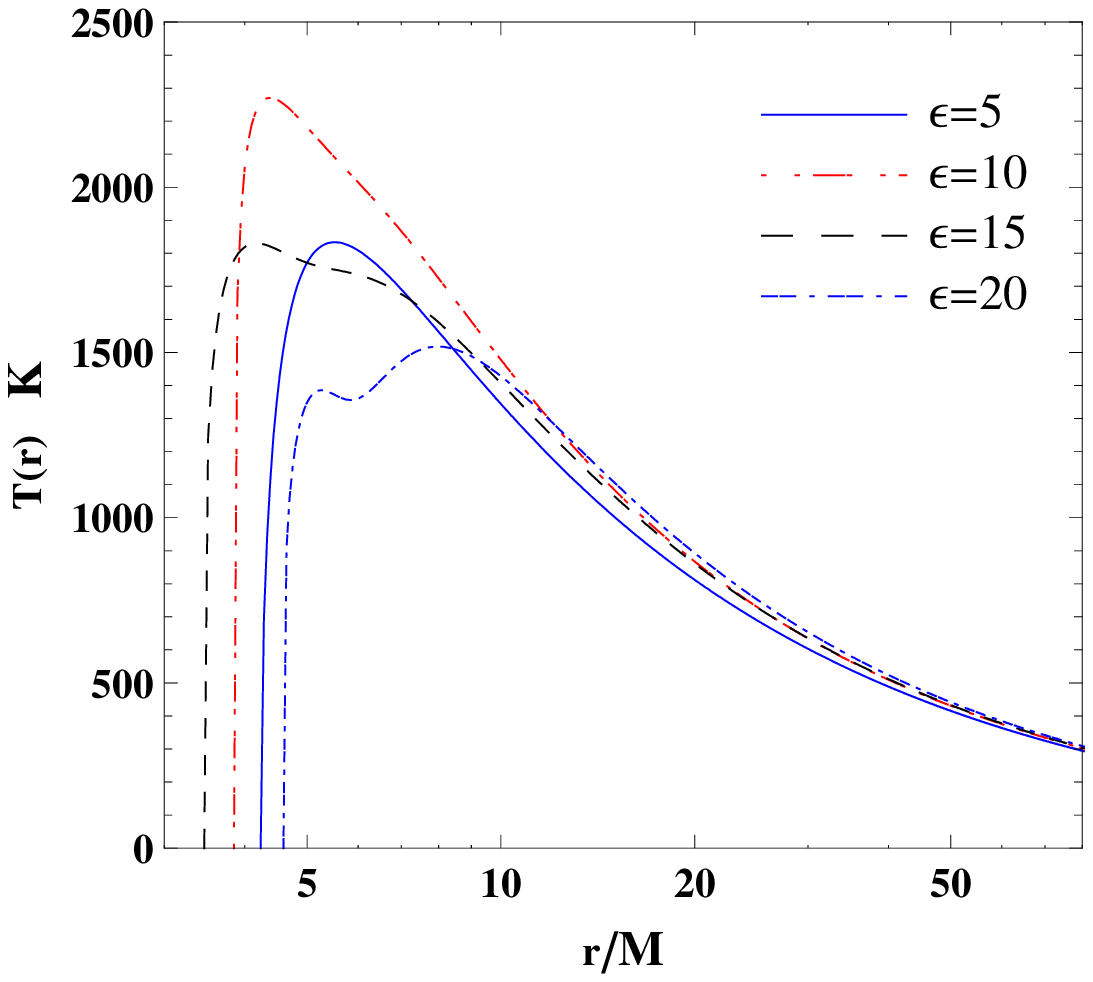}\includegraphics[width=5.4cm]{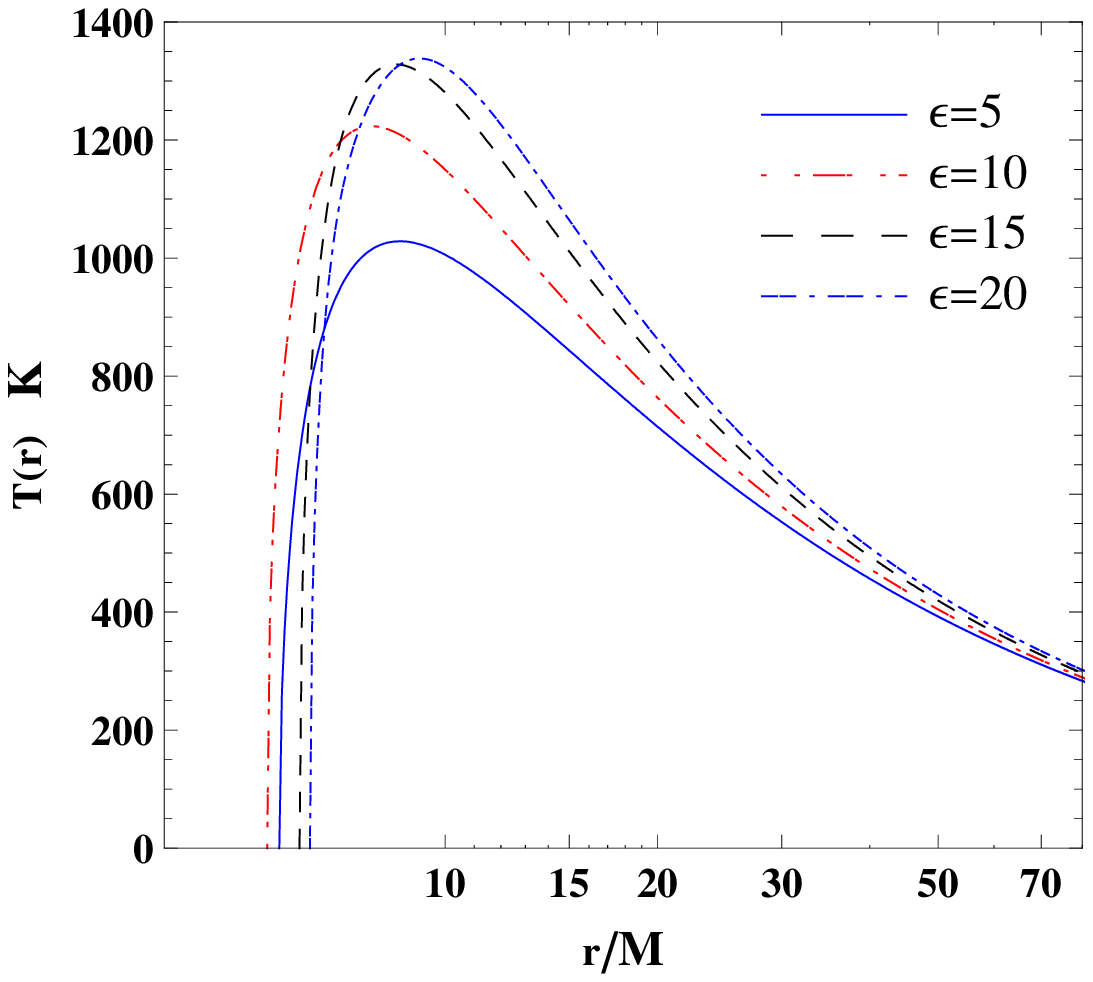}
\includegraphics[width=5.4cm]{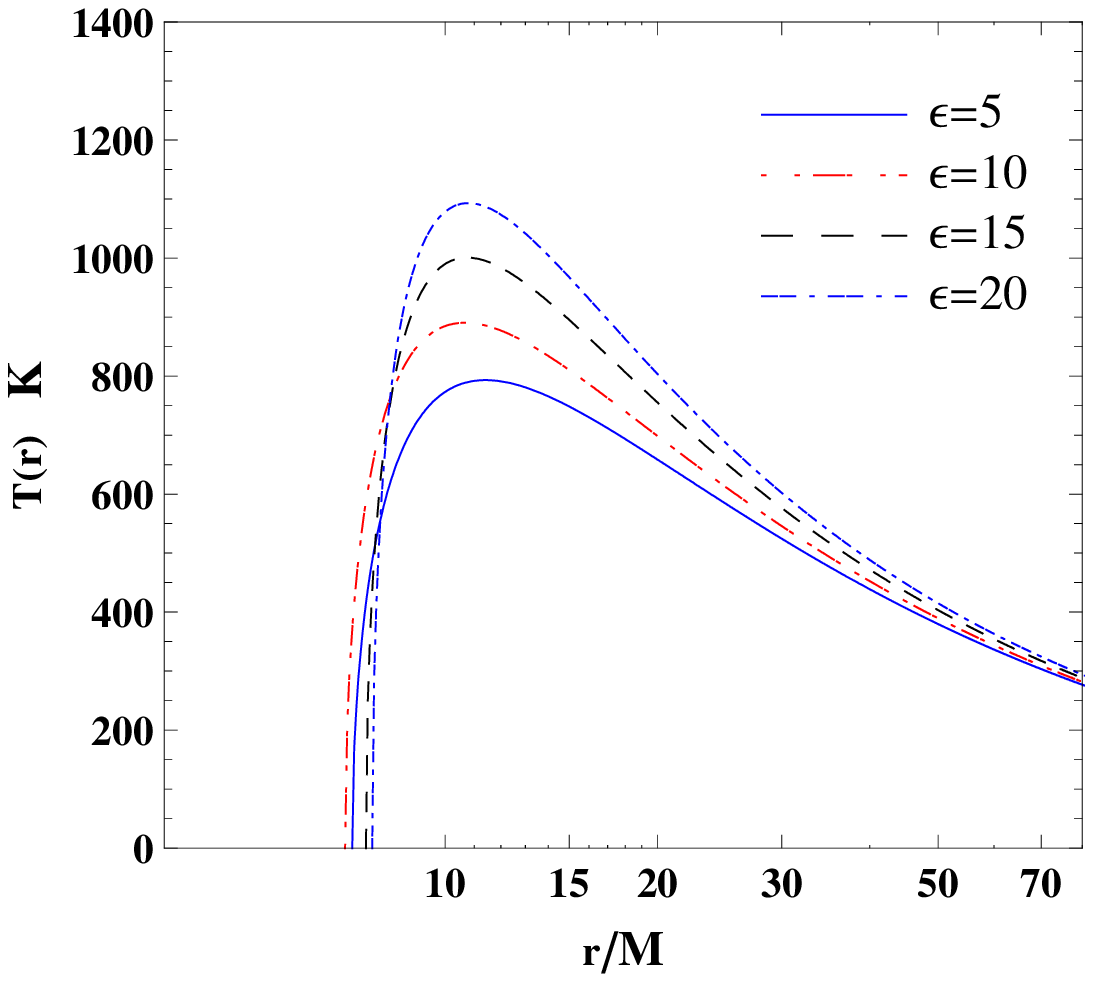}
\caption{Variety of the temperature $T$  with the deformed parameter
$\epsilon$ in the thin disk around the rotating non-Kerr black hole.
The two panels at the left, middle and right correspond to the cases
when $a=0.5$, $a=0$ and $a=-0.5$, respectively. Here, we set the
total mass of the black hole $M=10^6M_{\odot}$ and the mass
accretion rate $\dot{M_0}=10^{-12}M_{\odot}\;yr^{-1}$. }
\end{center}
\end{figure}

Let us now to probe the effects of $\epsilon$ on the radiation
temperature and the spectra of the disk around the rotating non-Kerr
black hole. In the steady-state thin disk model \cite{Page,Thorne},
it is assumed generally that the accreting matter is in
thermodynamic equilibrium, which means that the radiation emitted by
the disk surface can be considered as a perfect black body
radiation. The radiation temperature $T(r)$ of the disk is related
to the energy flux $F(r)$ through the expression
$T(r)=[F(r)/\sigma]^{1/4}$, where $\sigma$ is the Stefan-Boltzmann
constant. This means that the dependence of $T(r)$ on $\epsilon$ is
similar to that of the energy flux $F(r)$ on $\epsilon$, which is
also shown in Fig.(4).  Repeating the operations in \cite{Torres},
one can find that the observed luminosity $L(\nu)$ for the thin
accretion disk around the rotating non-Kerr black hole can be
expressed as
\begin{eqnarray}
L(\nu)=4\pi d^2I(\nu)=\frac{8\pi
h\cos{\gamma}}{c^2}\int^{r_f}_{r_{i}}\int^{2\pi}_{0}\frac{\nu_e^3\sqrt{-G}dr
d\phi}{e^{h\nu_e/KT(r)}-1},\label{emspe}
\end{eqnarray}
The emitted frequency is given by $\nu_e=\nu(1+z)$, where the
redshift factor can be written as
\begin{eqnarray}
1+z = \frac{1+\Omega r \sin\phi\sin\gamma}{\sqrt{-g_{tt}-2\Omega
g_{t\phi}-\Omega^2g_{\phi\phi}}},
\end{eqnarray}
where we have neglected the effect of light bending
\cite{CBa3,Luminet, Bhattacharyya}.
\begin{figure}[ht]
\begin{center}
\includegraphics[width=5.4cm]{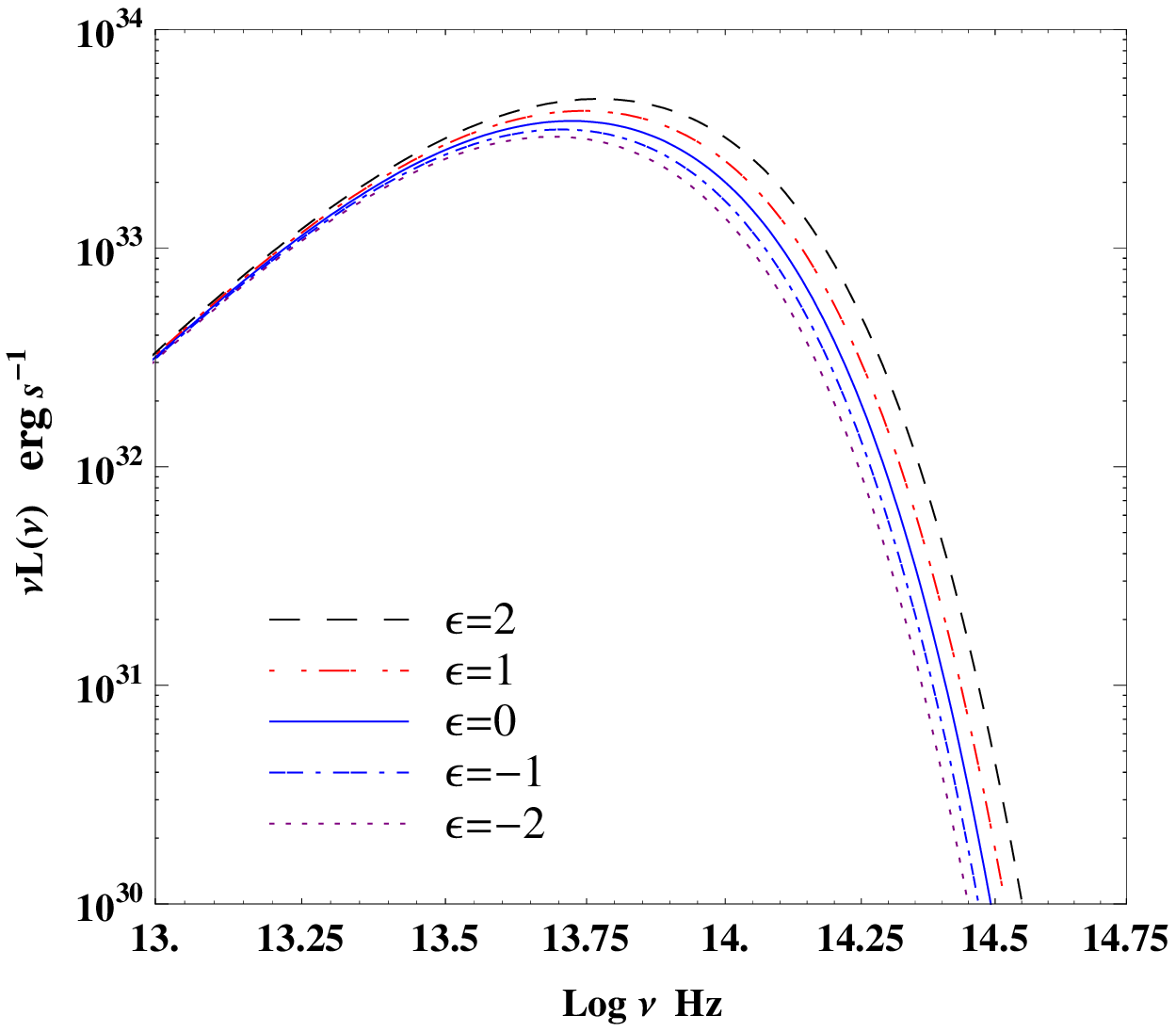}\includegraphics[width=5.4cm]{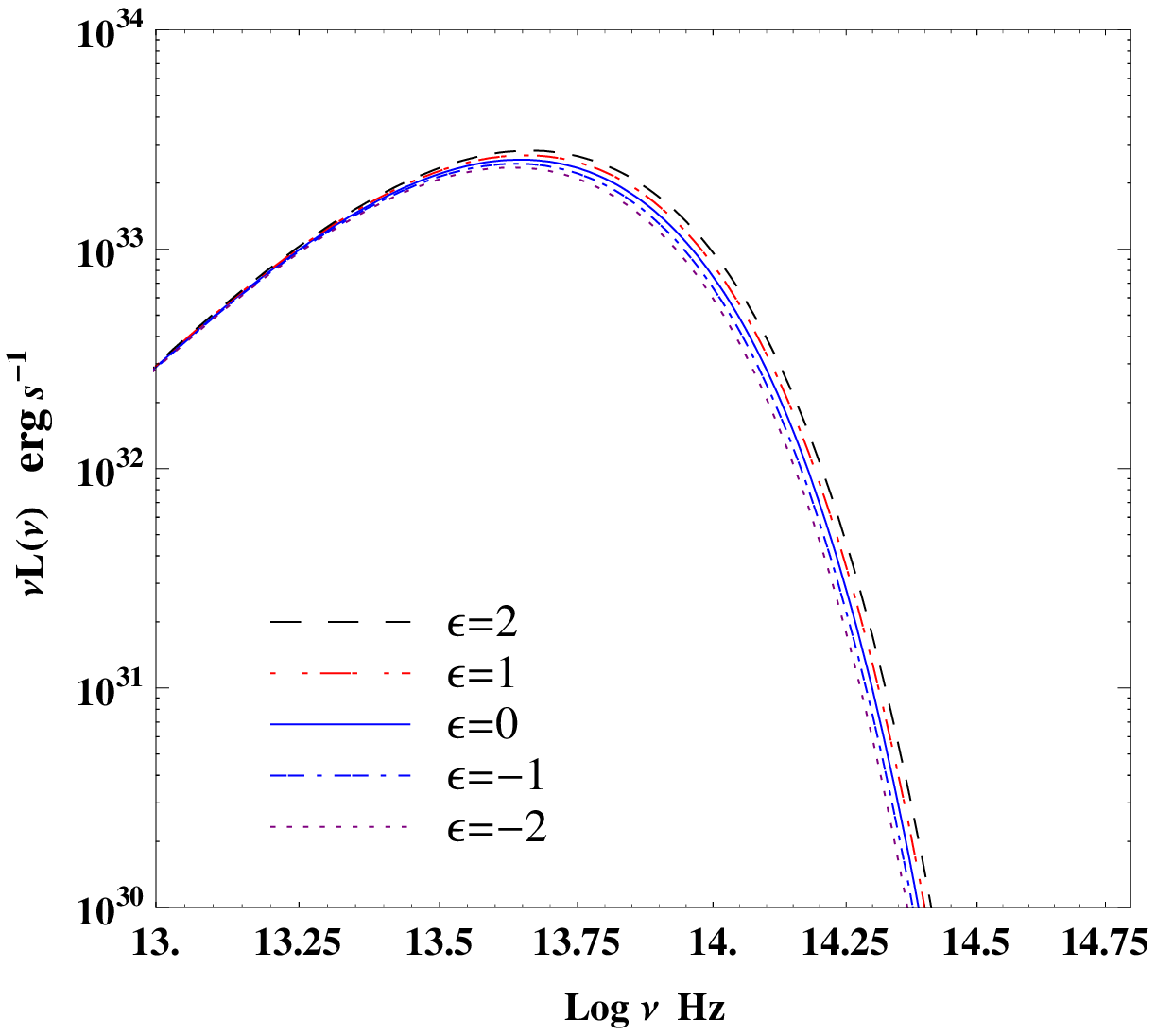}
\includegraphics[width=5.4cm]{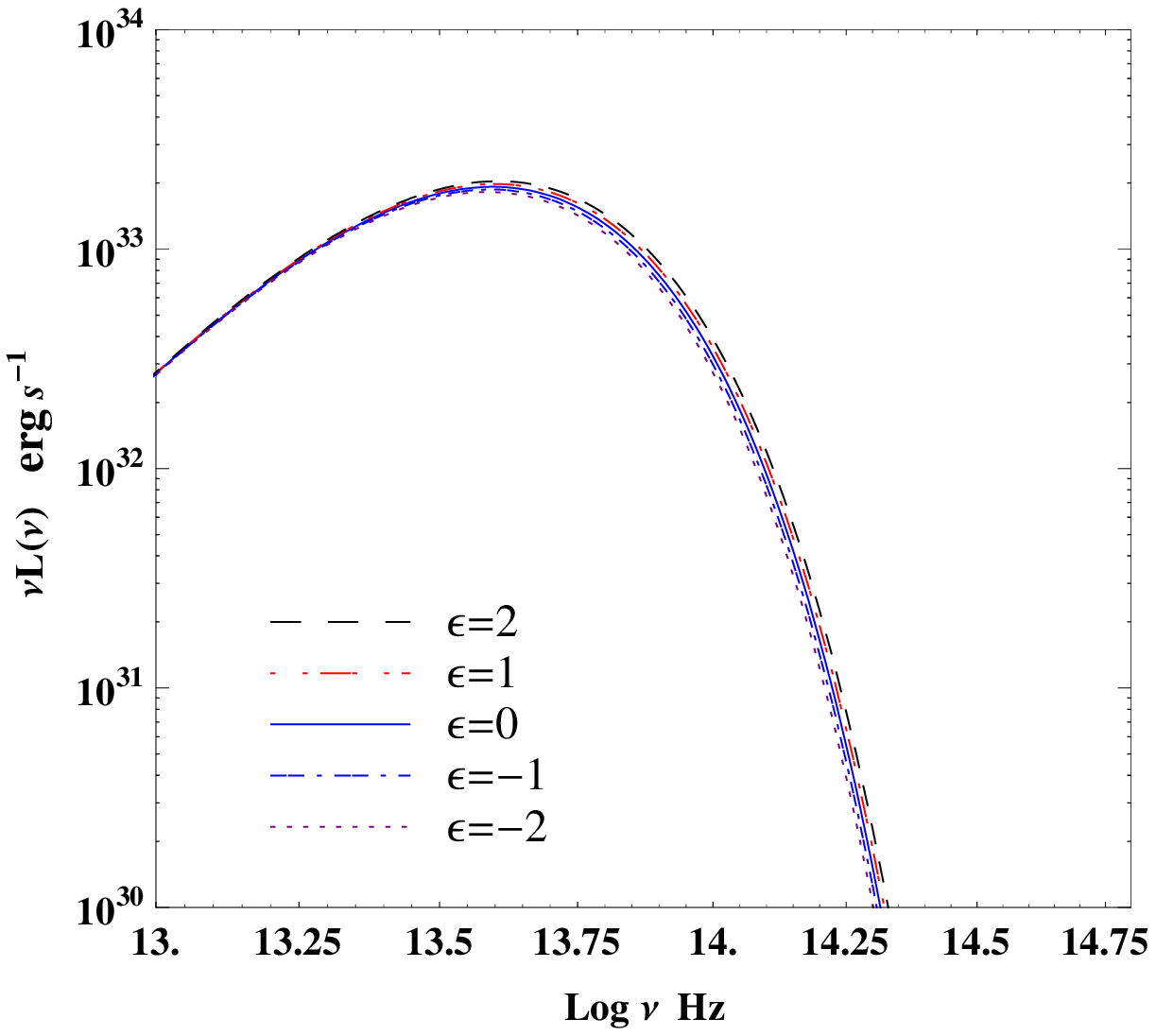}\\
\includegraphics[width=5.4cm]{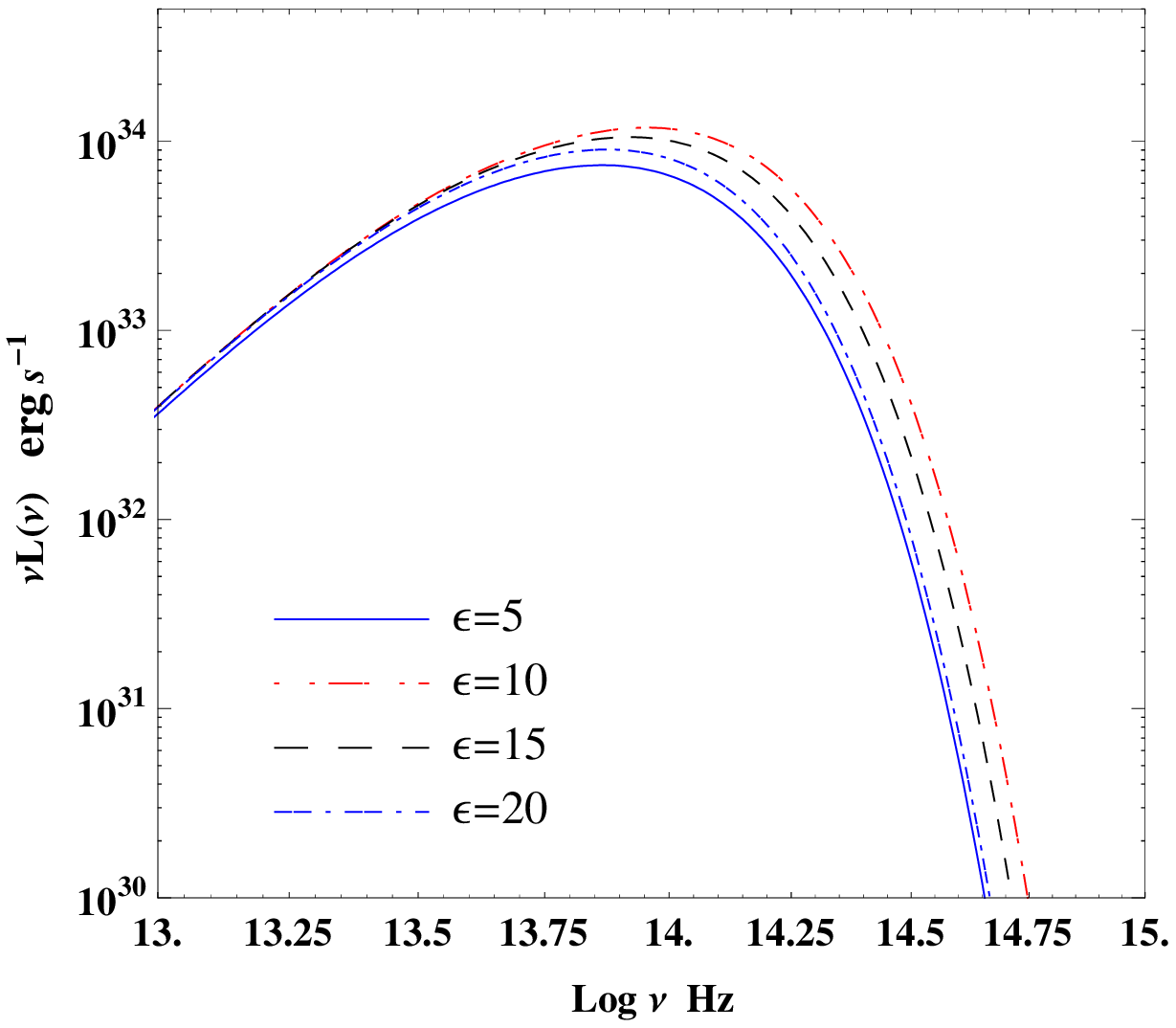}\includegraphics[width=5.4cm]{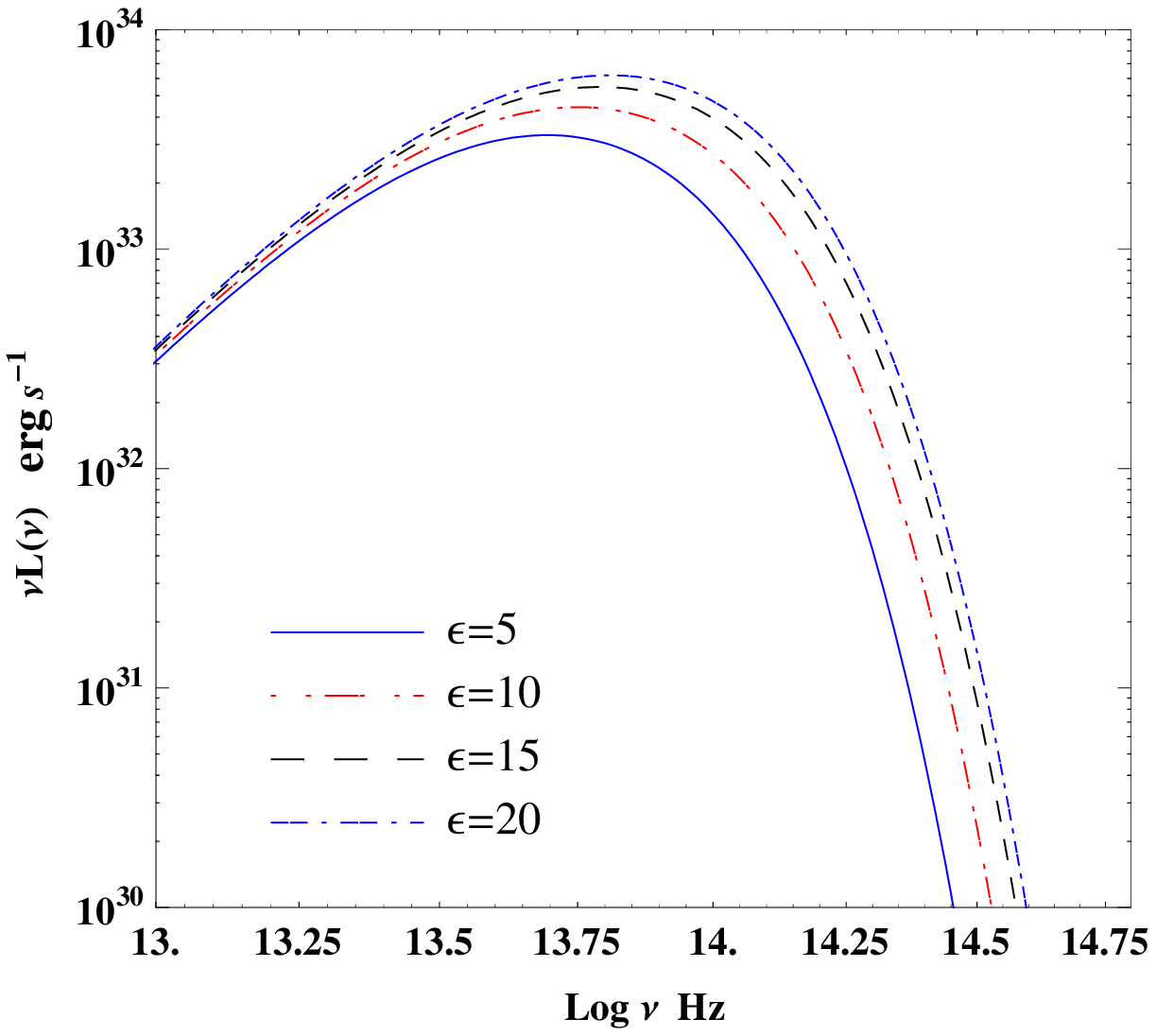}
\includegraphics[width=5.4cm]{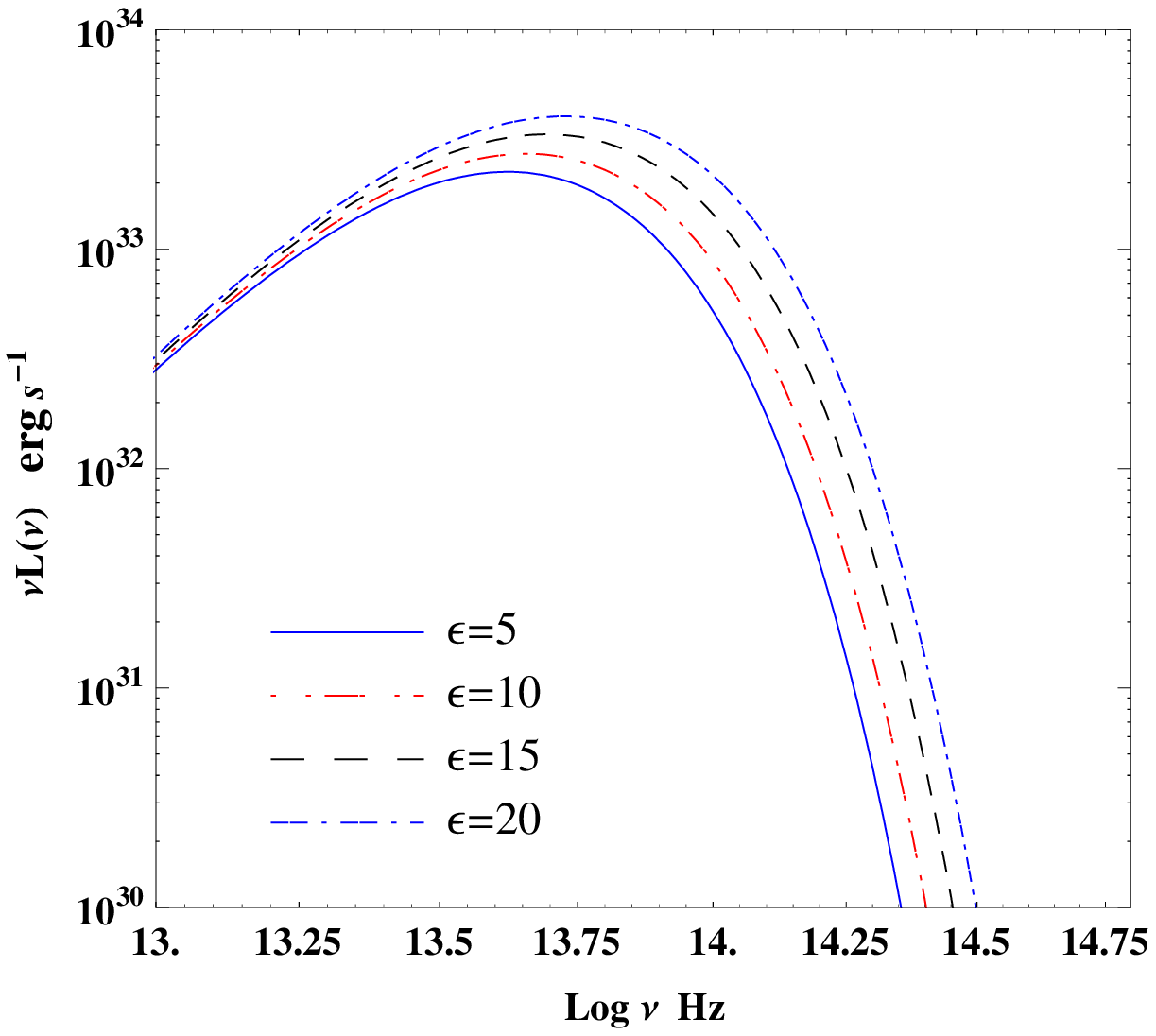}
\caption{Variety of the emission spectrum with the deformed
parameter $\epsilon$ in the thin disk around the rotating non-Kerr
black hole. The two panels at the left, middle and right correspond
to the cases when $a=0.5$, $a=0$ and $a=-0.5$, respectively. Here,
we set the total mass of the black hole $M=10^6M_{\odot}$, the mass
accretion rate $\dot{M_0}=10^{-12}M_{\odot}\;yr^{-1}$ and the disk
inclination angle $\gamma=0^{\circ}$.}
\end{center}
\end{figure}
The quantity $d$ is the distance to the source, $I(\nu)$ is the
thermal energy flux radiated by the disk, and  $\gamma$ is the disk
inclination angle. The quantities $r_f$ and $r_i$ are the outer and
inner border of the disk, respectively. In order to calculate the
luminosity $L(\nu)$ of the disk, we choose $r_i=r_{ms}$ and
$r_f\rightarrow \infty$ since the flux over the disk surface
vanishes at $r_f\rightarrow \infty$ in the rotating non-Kerr black
hole spacetime. Resorting to numerical method, we calculate the
integral (\ref{emspe}) and present the spectral energy distribution
of the disk radiation in Fig.(5). As $\epsilon<\epsilon_{c3}$ for
$a<0$ and $\epsilon<\epsilon_{c1}$ for $a>0$, we find that the
larger value of $\epsilon$ leads to the higher cut-off frequencies
and the higher observed luminosity of the disk. Moreover, we also
find that in this parameter regime  the effect of the deformed
parameter $\epsilon$ on the spectra becomes more distinct for the
prograde particles and more tiny for the retrograde ones. As
$\epsilon>\epsilon_{c3}$ for $a<0$ and $\epsilon>\epsilon_{c1}$ for
$a>0$, both the luminosity and cut-off frequencies decrease with
$\epsilon$. Like in the energy flux $F(r)$,  the effect of the
rotation parameter $a$ on disk spectra vanishes gradually with the
increase of $\epsilon$ in this region.

\section{summary}

A four-dimensional rotating black hole was proposed recently by
Johannsen \textit{et al} \cite{TJo} to test the no-hair theorem.
This black hole deviates from the usual Kerr black hole with a
deformed parameter and an unbound rotation parameter. The study of
such a rotating non-Kerr black hole can help us to understand more
deeply about the no-hair theorem and the cosmic censorship
conjecture in the strong gravity field regime. In this Letter, we
have studied the properties of the thin accretion disk in the
rotating non-Kerr black hole background. Our results show that the
deformed parameter $\epsilon$ imprints in the energy flux,
temperature distribution and emission spectra of the disk. For the
case in which the black hole is more oblate than a Kerr black hole,
the large deviation diminishes the energy flux, the conversion
efficiency, the radiation temperature, the spectra luminosity and
cut-off frequency of the thin accretion disk. For the black hole
with positive $\epsilon$, we find that these physical quantities of
disk increases with $\epsilon$ in the range of $a<0$ with
$\epsilon<\epsilon_{c3}$ and of $a>0$ with $\epsilon<\epsilon_{c1}$.
For the rapidly rotating black hole, the effect of the deformed
parameter $\epsilon$ on the physical quantities of the thin disk
becomes more distinct for the prograde particles and more tiny for
the retrograde ones. However, for the cases the parameters located
in the range of $a<0$ with $\epsilon>\epsilon_{c3}$ and of $a>0$
with $\epsilon>\epsilon_{c1}$, the situation is converse. The energy
flux, the conversion efficiency, the radiation temperature, the
spectra luminosity and cut-off frequency of the disk decrease with
$\epsilon$ and the effect originating from the rotation parameter
$a$ vanishes gradually with the increase of $\epsilon$. These
significant features,  at least in principle, may provide a
possibility to test the no-hair theorem in the future astronomical
observations.

\section{\bf Acknowledgments}

We would like to thank Prof. C. Bambi for his useful comments and
discussions. This work was  partially supported by the NCET under
Grant No.10-0165, the PCSIRT under Grant No. IRT0964 and the
construct program of key disciplines in Hunan Province. J. Jing's
work was partially supported by the National Natural Science
Foundation of China under Grant No.10935013; 973 Program Grant No.
2010CB833004.

\end{document}